\documentclass[amsmath,amssymb,pra]{revtex4}
\usepackage[cp1251]{inputenc}
\usepackage[english,russian]{babel}
\usepackage{color}
\usepackage{graphicx}
\usepackage{dcolumn}
\usepackage{braket}
\usepackage{amsmath}

\expandafter\let\csname equation*\endcsname\relax
\expandafter\let\csname end equation*\endcsname\relax

\usepackage{amsmath}

\begin{document}

\title{A PROTOCOL OF MEASUEREMENTS PROVIDING DIRECT, COMPLETE AND SINGLE-VALUED RECOVER OF ALL A-PRIORY UNKNOWN  PARAMETERS OF BIPHOTON POLARIZATION QUTRITS}

\author{M.V. FEDOROV$^{1,2}$,  C.C. MERN0VA, K.V. SLIPOROD\\
$^1$ Prokhorov General Physics Institute of the Russian Academy of Sciences\\
$^2$  Higher School of Economics, Moscow, Russia}

\begin{abstract}
We suggest and describe the protocol of measurements providing completely probabilistic representation of all parameters of biphoton polarization qutrits, i.e providing explicit expressions for all parameters of qutrits via the probabilities of getting those or other results in measurements.
\end{abstract}
\maketitle

\section{INTRODUCTION}

As known, spontaneous parametric down-connversi0n (SPDC) \cite{Klyshko, Harris, Magde} is one of the most often used methods for production of entangled biphoton states. In SPDC photons are born by pairs in anisotropic crystals under the action of the pump. The totality of such photon pairs form biphoton states which can be entangled. The pairs of variables of the SPDC photons are their frequencies, propagation angles outside the crystal, or polarizations of photons. Most often entanglement in respect to any of these pairs of variables can be considered separately from others. In this work we concentrate our attention on polarization biphoton states. In practice purely polarization biphoton states arise in the  case of collinear and frequency-degenerate regime of SPDC. As known well, there are only three basis polarization biphoton states, in which both photons in pairs are either horizontally or both vertically polarizes, and the third basis state with different orthogonal (horizontal and vertical) polarizations of two photons in pairs. Superposition of these basis states with coefficients $C_{1,3,2}$ is known as the polarization biphoton qutrit. Values of the parameters $C_{1,2,3}$ can be varied and controlled in the process of creation of these states. For this reason, in principle, these parameters can be used as a storage medium. On the other hand, the values of $C_{1,2,3}$ can be a-priory unknown to the observer, and for this reason one can need a procedure of measuring providing direct, complete and single-valued recovering of all unknown parameters of biphoton qutrits. As far as we know, for qutrits of a general form, description of such a procedure is missing in literature, and in this work we identify the set of measurements, which are sufficient for achieving the formulated goals.

Note that, in principle, the recovery of unknown parameters of quantum states belongs to the field of science known as quantum tomography The works in this field are numerous and only a small part of them to be mentioned here is \cite{Thew, Che, DM, L, Bogd, Bogd2}. However most often quantum tomography deals with density matrices of quantum states rather than with pure quantum states themselves. In contrast to this, in this work we do not consider density matrices of qutrits at all. We speak about direct recovery  of the qutrit's parameters as they appear in the qutrit's  state vectors or wave functions. In such formulation the problem of recovering the qutrit's parameters was partly considered in the work \cite{NJP}, though there the complete solution was not found. So, here we suggest and describe the scheme which consists of several steps finally gives a straight way for complete and single-valued recovery of all parameters of biphoton polarization qutrits.

Note also, that below we describe sequences of measurements. the results of which are probabilities of getting those or other outcomes from detectors counting photons. In this way parameters of qutrits appear to be expressed in terms of these probabilities. To some extent such results correlate with the ideas of V.I. Manko  on probabilistic formulation of quantum mechanics \cite{Manko}.

\section{Formulation of the problem}

Thus, as explained above and as it is well known, in a general case the state vector of the biphoton polarization qutrits  is given by
\begin{equation}
\label{QTR-HV}
\ket{\Psi} = C_1\ket{2_H}+C_2\ket{1_H,1_V}+C_3\ket{2_V}\equiv
\left(\frac{C_1}{\sqrt{2}}\, a_{H}^{\dag\,^2}+C_2 a_{H}^\dag a_{V}^\dag+\frac{C_3}{\sqrt{2}}\,a_{V}^{\dag\,^2}\right)\ket{0},
\end{equation}
where the numbers 1 and 2 in the ket- state vectors are the numbers of photons, $H$ and $V$ refer to their horizontal and vertical polarizations,  $a_H^\dag$ and $a_V^\dag$ are the single-photon creation operators, and $\ket{0}$ is the vacuum state vector; $C_i=|C_i|exp(i\varphi_i)$ are the  complex constants with the phases $\varphi_i$, and their absolute values obey the normalization condition
\begin{equation}
 \label{Norm}
 |C_1|^2+|C_2|^2+|C_3|^2=1.
\end{equation}
Besides, the common phase of the polarization qutrit is unmeasurable and can be chosen arbitrarily. Let us choose it below in such a way that the constant $C_2$ becomes real and positive. As the result, the problem of recovering all the unknown qutrit's parameters is reduced to the task of measuring four independent  constants: two absolute values $|C_1|$ and $|C_3|$ and two phases $\varphi_1$ and $\varphi_3$ of the constants $C_1$ and $C_3$. The first part of measuring $|C_1|$ and $|C_3|$ can be solved rather easily with the help of the routine coincidence-scheme measurements briefly described in the next section. As for the phases $\varphi_1$ and $\varphi_3$, the task of their measurement is  somewhat more tricky, and one of the ways of doing this is suggested and described in the subsequent sections.

\section{Measurement of  $|C_{1,2,3}|$ (Step 1).}

In all cases discussed here and in subsequent sections all measurements are assumed to be carried out in the frame of the coincidence schemes, the simplest of which is shown in Figure \ref{Figure1}
\begin{figure}[h]
  \label{Figure1}
 \centering
  \includegraphics[width=8cm]{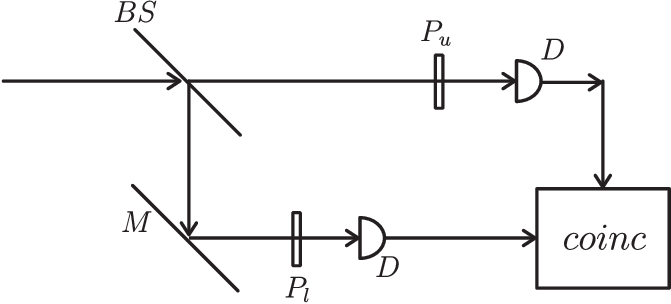}
  \caption{Coincidence scheme for measuring the absolute values of the qutrit's parameters $|C_{1,2,3}|$; $BS$ is the beamsplitter, $M$ denotes the mirror, $D$ are detectors, $P_u$ and $P_l$ are polarizers in the upper and lower channels.}\label{Figure1}
\end{figure}

At the beamsplitter each photon pair with the probability 50\% can be either split between two channels or  move unsplit  either straight or down. The coincidence scheme registers only split pairs of photons. There are four possibilities to install the polarizers  $P_u$ and $P_l$: 1) $P_u=P_l=P_H$, 2) $P_u=P_l=P_V$, 3) $P_u=P_H,\; P_l=P_V$, and  4) $P_u=P_V,\; P_l=P_H$. In the first two cases detectors count only split pairs of photons with coinciding polarizations $HH$ or $VV$, and let the numbers of counts in these cases be $N_{HH}$ and $N_{VV}$. In the cases 3 and 4 detectors will count only split $H_uV_l$ and $H_lV_u$ pairs and let the numbers of these counts be $N_{HV}$ and $N_{VH}$. In all four cases the measurement times must be identical. Then the sum of all count numbers is
\begin{equation}
 \label{total}
 N_{tot}=N_{HH}+N_{VV}+N_{HV}+N_{VH},
\end{equation}
and in can be used for the probabilities, with which all four kinds of photon pairs are presented in the qutrit (\ref{QTR-HV})
\begin{equation}
 \label{prob}
 w_{HH}=\frac{N_{HH}}{N_{tot}}=|C_1|^2,\quad
 w_{HV}^{tot}=w_{HV}+w_{VH}=2w_{HV}=\frac{N_{HV}+N_{VH}}{N_{tot}}=|C_2|^2,\quad
 w_{HH}=\frac{N_{HH}}{N_{tot}}=|C_3|^2
 \end{equation}
and of course
\begin{equation}
 \label{Norm}
 w_{HH}+w_{HV}^{tot}+w_{HH}=|C_1|^2+|C_2|^2+|C_3|^2=1
 \end{equation}
Equations (\ref{prob}) establish relations between the absolute values of all three qutrit's parameters $|C_{1,2,3}|$ and the results of measurements in the scheme of Figure 1, and in the same time these equations provide fully probabilistic representation of $|C_{1,2,3}|$. The next steps concern determination of the phases $\varphi_1$ and $\varphi_3$ of the constants $C_1$ and $C_3$ at $C_2$ known and real ($\varphi_2=0$).

\section{Measurements with polarizers turned for 45$^\circ$ (Step 2)}

With the optical axes of both polarizers in Figure \ref{Figure1} turned  for 45$^\circ$ around the $z-$ axis,
 the polarizers $P_u$ и $P_l$ turn into $P_{45^\circ}$ and they select only photons, polarization of which is directed along the directions at 45$^\circ$ or at 135$^\circ$ with respect to the horizontal one (0$^\circ$). To learn what can be registered in the coincidence scheme now, we have to transform the general expression for the qutrit's state vector (\ref{QTR-HV}) from the $(H,V)$ basis to the basis ${45^\circ,135^\circ}$, e.g. with the help of the transformation formulae for the photon's creation operators
\begin{equation}
\label{transf1}
a_H^\dag=\frac{1}{\sqrt{2}}(a_{45^\circ}^\dag-a_{135^\circ}^\dag),^\circ
\quad a_V^\dag=\frac{1}{\sqrt{2}}(a_{45^\circ}^\dag+a_{135}^\dag).
\end{equation}
As a result, the transformed expression for state vector can be reduced to the same form as the expression ((\ref{QTR-HV}))
\begin{equation}
\label{qutr-45-135}
\ket{\Psi}=\left(\frac{B_1}{\sqrt{2}}a^{\dag\,2}_{45^\circ}+B_2 a^\dag_{45^\circ}a^\dag_{135^\circ}+\frac{B_3}{\sqrt{2}}a^{\dag\,2}_{135^\circ}\right)\ket{0}=
B_1\ket{2_{45^\circ}}+B_2\ket{1_{45^\circ},1_{135^\circ}}+B_3\ket{2_{135^\circ}},
\end{equation}
but with different coefficients $B_{1,2,3}$ given by
\begin{equation}
\label{B-1-2-3}
B_1=\frac{C_1+C_3}{2}+\frac{C_2}{\sqrt{2}},\quad B_2=\frac{-C_1+C_3}{\sqrt{2}},\quad B_3=\frac{C_1+C_3}{2}
-\frac{C_2}{\sqrt{2}}.
\end{equation}

Similarly to the case of $P_H$ и $P_V$ polarizers and $H-V$ basis, the parameters $B_{1,2,3}$ are related directly with the set of experimentally measurable probabilities of registering pairs of photons with various polarizations in the $45^\circ-135^\circ$ basis
\begin{equation}
\label{prob-45-135}
 |B_1|^2=w_{45^\circ,45^\circ},\quad  |B_2|^2=w^{tot}_{45^\circ,135^\circ}=2w_{45^\circ,135^\circ}=2w_{135^\circ,45^\circ},
 \quad{\rm and}\quad
 |B_3|^2=w_{135^\circ,135^\circ}.
 \end{equation}

These equation and the transformation formulas od Equation (\ref{B-1-2-3}) can be used to get the following two equations for the phases $\varphi_1$ и $\varphi_3$
\begin{equation}
 \label{eq-for-cos-diff}
 2w_{45^\circ,135^\circ}=|B_2|^2=\frac{|C_1-C_3|^2}{2}=\frac{w_{HH}+w_{VV}}{2}-
 \sqrt{w_{HH}w_{VV}}\,\cos(\varphi_1-\varphi_3)
 \end{equation}
and
\begin{equation}
 \label{diff-of-prob}
 w_{45^\circ,45^\circ}-w_{135^\circ,135^\circ}=|B_1|^2-|B_3|^2=\sqrt{2}\,C_2\,Re(C_1+C_3)=
 2\sqrt{w_{HV}}\big(\sqrt{w_{HH}}\cos\varphi_1+\sqrt{w_{VV}}\cos\varphi_3\big)
\end{equation}

But, clearly enough, these equations are insufficient for the single-valued definition of phases $\varphi_{1,3}$. In particular, because they contain only cosine dependencies on the phases $vaphi_1$, $vaphi_3$ and on their difference, which inevitably makes their solutions non-single-valued. Additional equation are needed and their derivation is discussed in the next sections, which is preceded by a brief reminder on the Jones matrices and features of phase wave plates.

\section{Polarization wave functions of biphoton qutrits and their transformations}

In addition to the description of single-photon states and qutrits in terms of their state vectors, the same states can be characteriz4d by their wave functions. In particular, the wave functions of the states $\ket{1_H}=a_H^\dag\ket{0}$ and  $\ket{1_V}=a_V^\dag\ket{0}$ in the matrix representation have the form
$\left(
\begin{matrix}
 1\\
 0
 \end{matrix}
 \right )$
with the upper and lower lines corresponding to the horizontal and vertical polarizations. The wave functions of biphoton basis polarization states
 $\ket{2_H}$, $\ket{2_V}$ и $\ket{1_H,1_V}$ are given by
 \begin{equation}
 \label{wf-baz-st}
 \psi_{HH}=\left(
\begin{matrix}
 1\\
 0
 \end{matrix}
 \right)_1
 \left(
\begin{matrix}
 1\\
 0
 \end{matrix}
 \right)_2,
 \quad
  \psi_{VV}=\left(
\begin{matrix}
 0\\
 1
 \end{matrix}
 \right)_1
 \left(
\begin{matrix}
 0\\
 1
 \end{matrix}
 \right)_2,
 \quad
 \psi_{HV}=\frac{1}{\sqrt{2}}
\left[ \left(\begin{matrix}
 1\\
 0
 \end{matrix}
 \right)_1
 \left(\begin{matrix}
 0\\
 1
 \end{matrix}
 \right)_2+\left(\begin{matrix}
 0\\
 1
 \end{matrix}
 \right)_1
 \left(\begin{matrix}
 1\\
 0
 \end{matrix}
 \right)_2
 \right],
\end{equation}
where the indices 1 and 2 are the numbers of two indistinguishable photons. Superposition of these basis wave function gives the wave function of the polarization qutrit of a general form\begin{equation}
  \label{WF+QTR}
  \Psi_{QTR}=C_1\psi_{HH}+C_2\psi_{HV}+C_3\psi_{VV}.
 \end{equation}
 Transformation of polarization wave functions by various phase plates is known to be provided by the Jones matrices \cite{Jones,Fowles,Che2}, $M(\alpha,\,\varphi)$. The parameters on which the Jones matrices depend are the angle $\alpha$ between the optical axis and the horizontal direction and the additional phase shift $\varphi$ in the wave function of the vertically polarized photon provided by the phase plate with $\alpha$.
 A general form of the Jones matrix is
\begin{equation}
 \label{Jones}
 M(\alpha,\varphi)=
 \left(
 \begin{matrix}
 \cos^2\alpha + e^{i\varphi}\sin^2\alpha\quad
 (1 - e^{i\varphi})\sin\alpha\cos\alpha\\
 (1 - e^{i\varphi})\sin\alpha\cos\alpha\quad e^{i\varphi}\cos^2\alpha + \sin^2\alpha
 \end{matrix}
 \right)
\end{equation}
The following equation characterizes action of the general-form phase plate on the wave functions of the horizontally and vertically polarized single-photon states
\begin{equation}
\label{transf-single}
M\left(\begin{matrix}1\\0\end{matrix}\right)= \left(\begin{matrix}\cos^2\alpha + e^{i\varphi}\sin^2\alpha\\(1 - e^{i\varphi})\sin\alpha\cos\alpha\end{matrix}\right),\quad
M\left(\begin{matrix}0\\1\end{matrix}\right)= \left(\begin{matrix}(1 - e^{i\varphi})\sin\alpha\cos\alpha\\e^{i\varphi}\cos^2\alpha + \sin^2\alpha\end{matrix}\right)
\end{equation}
In the case of the phase-plate transformation of the biphoton wave functions, the transformation rules (\ref{transf-single}) must be applied to each of two terms in products of two columns in Equations (\ref{wf-baz-st}).

The most often used phase plates are the halph-wavelength ($\varphi=\pi$) and the quarter-wavelength ($\varphi=\pi$)
plates, and their Jones matrices are given by
\begin{equation}
 \label{Jones-halph}
 M(\alpha,\pi)=\left(\begin{matrix}
 \cos 2\alpha\;\;\;\sin 2\alpha\\
 \;\sin 2\alpha\;-\cos 2\alpha\end{matrix}\right),
 \quad
 M(\alpha,\pi/2)=\left(\begin{matrix}
 \cos^2\alpha+i\sin^2\alpha\;\;\;\;(1-i)\sin\alpha\cos\alpha\\
 (1-i)\sin\alpha\cos\alpha\;\;\;i\cos^2\alpha+\sin^2\alpha\end{matrix}\right).
 \end{equation}

It's an open question whether it's possible or not to find a combinations of transformations by these phase plates for getting appropriate equations additional to (\ref{eq-for-cos-diff}) and (\ref{diff-of-prob}) and providing complete and single-valued recovering of all qutrit's parameters (i.e. the phases $\varphi_1$ and $\varphi_3$). But below we use a different approach, the key element of which is the use of the $\lambda/8$ phase plate. The Jones matrix of such phase plate is given by
\begin{equation}
 \label{lambda/8}
 M(\alpha,\pi/4)=\left(\begin{matrix}
 \cos^2\alpha+\sqrt{i}\,\sin^2\alpha\;\;\;\;\left(1-\sqrt{i}\,\right)\sin\alpha\cos\alpha\\
 \left(1-\sqrt{i}\,\right)\sin\alpha\cos\alpha\;\;\;\sqrt{i}\,\cos^2\alpha+\sin^2\alpha\end{matrix}\right).
\end{equation}

\section{Preliminary transformation of the biphoton qutrit by the $\lambda/8$ phase plate with $\alpha=0$ (Step 3).}

The scheme we suggest at this stage and shown in Figure 2 is similar to that considered above and shown in Figure 1 with only one change: the original biphoton qutrit is assumed to be preliminary modified by the phase plate $lambda/8$ with horizontally directed optical axis.

\begin{figure}[h]
  \centering
  \includegraphics[width=8cm]{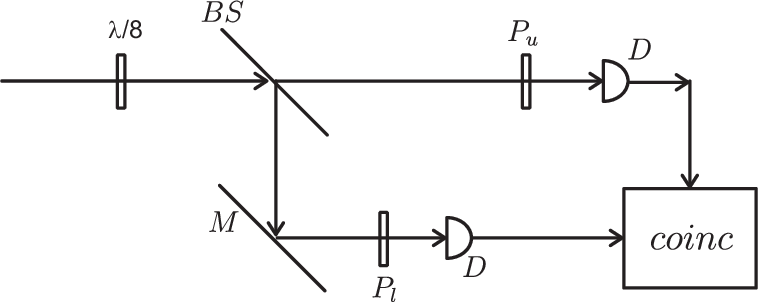}
  \caption{A scheme of observation with the qutrit preliminary modified by the $\lambda/8 $ phase plate with horizontally oriented optical axis.}\label{Figure2}
\end{figure}

The Jones matrix is in this case very simple
\begin{equation}
 \label{Jones lambd div 8}
 M(0,\pi/4)=\left(\begin{matrix}1\;\; 0\\0\;\; \sqrt{i}\end{matrix}\right).
\end{equation}
Transformation of single-photon states by such matrix is reduced simply to multiplication of one of them by $\sqrt{i}$: $M\left(\begin{matrix}1\\0\end{matrix}\right)=\left(\begin{matrix}1\\0\end{matrix}\right)$ и $M\left(\begin{matrix}0\\1\end{matrix}\right)=\sqrt{i}\left(\begin{matrix}0\\1\end{matrix}\right)$, and this results in changes in phases of coefficients in the qutrit:
\begin{equation}
 \label{Ctransf}
 C_1^{\lambda/8}=C_1,\;C_2^{\lambda/8}=\sqrt{i}\,C_2,; C_3^{\lambda/8}=i\,C_3.
\end{equation}

Like in Section 4, let us consider here only the case of polarizers $P_u$ and $P_v$turned for  45$^\circ$ from the $(0^\circ, 90^\circ)$ orientation. In the basis $45^\circ-135^\circ$ the preliminary modified qutrit's state vector takes the form
\begin{equation}
\label{qutr-45-135}
\ket{\Psi}=B_1^{\lambda/8}\ket{2_{45^\circ}}+B_2^{\lambda/8}\ket{1_{45^\circ},1_{135^\circ}}+B_3^{\lambda/8}\ket{2_{135^\circ}},
\end{equation}
where
\begin{equation}
 B_1^{\lambda/8}=\frac{C_1+i\,C_3}{2}+\sqrt{i}\,\frac{C_2}{\sqrt{2}},
 \quad
  \label{B-1-2-3-lambda/8}
  B_2^{\lambda/8}=\frac{-C_1+i\,C_3}{\sqrt{2}}\quad {\rm и} \quad
   B_3^{\lambda/8}=\frac{C_1+i\,C_3}{2}-\sqrt{i}\,\frac{C_2}{\sqrt{2}}.
\end{equation}
The squared absolute values of $B_1^{\lambda/8}$, $B_2^{\lambda/8}$, and $B_3^{\lambda/8}$ determine the observable probabilities of registration in the $(45^\circ-135^\circ)$ basis $(45^\circ-135^\circ)$
\begin{equation}
 \label{prob-transf}
 w^{\lambda/8}_{45^\circ,45^\circ}=|B_1^{\lambda/8}|^2,\quad
 w^{\lambda/8}_{45^\circ,135^\circ}=w^{\lambda/8}_{135^\circ,45^\circ}
 =\frac{|B_2^{\lambda/8}|^2}{2}, \quad w^{\lambda/8}_{135^\circ,135^\circ}=|B_3^{\lambda/8}|^2.
\end{equation}
With $B_2^{\lambda/8}$ taken from Equation (\ref{B-1-2-3-lambda/8}) the second of three Equations (\ref{prob-transf}) yields
\begin{equation}
 \label{eq-for-sin-diff-l/8}
 2w_{45,135}^{\lambda/8}
 =|B_2^{\lambda/8}|^2=\frac{|-C_1+iC_3|^2}{2}=
 \frac{1}{2}\Big[w_{HH}+w_{VV}-
 2\sqrt{w_{HH}w_{VV}}\sin(\varphi_1-\varphi_3)\Big].
\end{equation}

Jointly, the pair of equations (\ref{eq-for-cos-diff}) and (\ref{eq-for-sin-diff-l/8}) is sufficient for the single-valued recovery of the difference of phases $\delta=\varphi_1-\varphi_3$
\begin{equation}
 \label{cos-delta}
 \left\{
 \begin{matrix}
\displaystyle
 \cos\delta=\frac{w_{HH}+w_{VV}-4w_{45^\circ,135^\circ}}{2\sqrt{w_{HH}w_{VV}}}\equiv A_{\cos}\\
 \,\\
 \displaystyle
 \sin\delta=\frac{w_{HH}+w_{VV}-4w^{\lambda/8}_{45^\circ,135^\circ}}{2\sqrt{w_{HH}w_{VV}}}\equiv A_{\sin}
 \end{matrix}
 \right.
\end{equation}

If  $\delta\in[-\pi,\pi]$ the solution of Equations (\ref{cos-delta}) can be written explicitly as
\begin{equation}
 \label{arccos-delta}
 \delta=\arccos(A_{\rm cos})\times
 {\rm sign}(A_{\rm sin}).
 \end{equation}

The picture of Figure \ref{Figiure3} show how the solution of Equations (\ref{cos-delta}) can be found graphically at coinciding and different signs of the measured values $A_{\cos}$ and $A_{\sin}$ of Equations (\ref{cos-delta}).
 \begin{figure}[h]
  \centering
  \includegraphics[width=15cm]{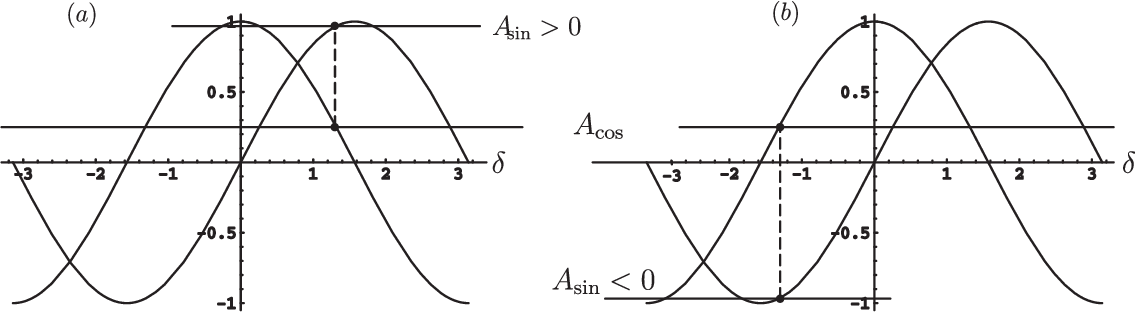}
  \caption{The decisions of Equations (\ref{cos-delta}) are shown by the dashed lines. }\label{fig3}
\end{figure}

Note that in a special case  of $C_2=0$ the results of this section finalize the procedure of recovering all qutrit's parameters. The reason is tat at $C_2=0$ the unmeasurable phase of the qutrit as  a whole can be chosen, e.g., in a way zeroing the phase $\varphi_3$, and then $\varphi_1=\delta$ where $\delta$ is known (\ref{arccos-delta}).

If however $C_2\neq 0$, measurement of $\delta$ is insufficient for completing recovering of all qutrit's parameters. The phases  $\varphi_1$ and $\varphi_3$ must be  found separately. For doing this, it is most convenient to use equation (\ref{diff-of-prob}) together with the result of an additional measurement with the optical axis of the $\lambda/8$ phase plate turned for $45^\circ$, as shown in the next section.

\section{Measurements with the optical axis of the  $\lambda/8$-phase plate turned for $45^\circ$ (Step 4).}

In the case $\alpha=\varphi=\pi/4$ the general expression (\ref{Jones} for the Jones matrix takes the form
\begin{equation}
 \label{Jonse45-45}
 M(\pi/4,\pi/4)=\left(\begin{matrix}
 \frac{1+\sqrt{i}}{2}\quad \frac{1 - \sqrt{i}}{2}\\
 \frac{1 - \sqrt{i}}{2}\quad\frac{1+\sqrt{i}}{2}
 \end{matrix}\right),
\end{equation}
and this matrix implements the following transformation of the single-photon polarization wave functions
\begin{equation}
\label{tr-single 4545}
M\left(\begin{matrix}1\\0\end{matrix}\right)=
\frac{1}{2}\left(\begin{matrix}1+\sqrt{i}\,\\1-\sqrt{i}\,\end{matrix}\right),\quad
M\left(\begin{matrix}0\\1\end{matrix}\right)=
\frac{1}{2}\left(\begin{matrix} 1-\sqrt{i}\, \\1+\sqrt{i}\,\end{matrix}\right),
\end{equation}
and the following transformation of the basis biphoton state vectors
\begin{gather}
 \nonumber
 M\ket{2_H}=\frac{(1+\sqrt{i}\,)^2}{4}\ket{2_H} +\frac{1-i}{2\sqrt{2}}\ket{1_H,1_V}+\frac{(1-\sqrt{i}\,)^2}{4}\ket{2_V},\\
 \label{Bas st-45}
 M\ket{1_H,1_V}=\frac{1-i}{2\sqrt{2}}\ket{2_H} +\frac{1+i}{2}\ket{1_H,1_V}+\frac{1-i}{2\sqrt{2}}\ket{2_V},\quad\quad\;\;\;\\
 \nonumber
 M\ket{2_V}=\frac{(1-\sqrt{i}\,)^2}{4}\ket{2_H} +\frac{1-i}{2\sqrt{2}}\ket{1_H,1_V}+\frac{(1+\sqrt{i}\,)^2}{4}\ket{2_V}.
\end{gather}
If the transformed qutrit's state vector is presented in the standard form
 \begin{equation}
\label{qutr-45}
\ket{\Psi}^{\pi/4}=B_1^{\lambda/8,\pi/4}\ket{2_H}+B_2^{\lambda/8,\pi/4}\ket{1_H,1_V}+B_3^{\lambda/8,\pi/4}\ket{2_V},
\end{equation}
then the expansion coefficients $B_{1,2,3}^{\lambda/8,\pi/4}$ can be easily found to be given by
\begin{gather}
\nonumber
B_1^{\lambda/8,\pi/4}=\frac{(1+\sqrt{i}\,)^2}{4}\,
C_1+\frac{1-i}{2\sqrt{2}}\,C_2+\frac{(1-\sqrt{i}\,)^2}{4}\, C_3,\\
\label{C-45}
B_2^{\lambda/8,\pi/4}=\frac{1-i}{2\sqrt{2}}\,(C_1+C_3)+\frac{1+i}{2}\,C_2,\quad\quad\;\;\;\\
\nonumber
B_3^{\lambda/8,\pi/4}=\frac{(1-\sqrt{i}\,)^2}{4}\, C_1+\frac{1-i}{2\sqrt{2}}\,C_2+\frac{(1+\sqrt{i}\,)^2}{4}\, C_3,
\end{gather}
where, as previously, $C_{1,3}=|C_{1,3}|\exp(i\varphi_{1,3})$, $Im(C_2)=0$ and $C_2>0$.

The middle of three Equations (\ref{qutr-45}) yields
\begin{equation}
 \label{abd-val}
 \left|B_2^{\lambda/8,\pi/4}\right|^2=\frac{1}{2}\,C_2^2+\frac{1}{4}|C_1+C_3|^2
 +\frac{1}{\sqrt{2}}\,C_2\,Re[-i(C_1+C_3)],
\end{equation}
where $C_1=\sqrt{w_{HH}}e^{i\varphi_1}$, $C_3=\sqrt{w_{VV}}e^{i\varphi_3}$ and $C_2=\sqrt{2w_{HV}}$; the probabilities  $w_{HH}$,$w_{VV}$, $w_{HV}$, as well as $\delta=\varphi_1-\varphi_3$ are assumed to be known from measurements described in previous sections. As the result, Equation (\ref{abd-val}) takes the form
\begin{gather}
  \nonumber
  w_{HV}^{\lambda/8,\pi/4}=\frac{1}{2}\,\left|B_2^{\lambda/8,\pi/4}\right|^2=\frac{w_{HV}}{2}+
 \frac{1}{8}\big(w_{HH}+w_{VV}+2\sqrt{w_{HH}w_{VV}}\,\cos\delta\big)+\\
  \label{HV-45}
  \frac{1}{2}\sqrt{w_{HV}}\,(\sqrt{w_{HH}}\sin\varphi_1+\sqrt{w_{VV}}\sin\varphi_3).
\end{gather}
As $\varphi_3=\varphi_1+\delta$, Equation (\ref{diff-of-prob}) together with Equation (\ref{HV-45}) can be rewritten as a pair of equations for two unknown values, $\sin\varphi_1$ and $\cos\varphi_1$:
\begin{equation}
 \label{modif}
\left\{\begin{matrix}
\Big(\sqrt{w_{HH}}+\sqrt{w_{VV}}\cos\delta\Big)\cos\varphi_1+\sqrt{w_{VV}}\sin\delta\sin\varphi_1=F,\\
\\
-\sqrt{w_{VV}}\sin\delta\cos\varphi_1+\Big(\sqrt{w_{HH}}+\sqrt{w_{VV}}\cos\delta\Big)\sin\varphi_1=F^{\lambda/8,\pi/4},
\end{matrix}\right.
\end{equation}
with
\begin{gather}
 \label{F}
 F= \frac{w_{45^\circ,45^\circ}-w_{135^\circ,135^\circ}}{2\sqrt{w_{HV}}}, \\
\label{F-45}
 F^{\lambda/8,\pi/4}=\frac{2w_{HV}^{\lambda/8,\pi/4}}{\sqrt{w_{HV}}}-\sqrt{w_{HV}}-
 \frac{1}{4\sqrt{w_{HV}}}\Big(w_{HH}+w_{VV}+\sqrt{w_{HH}w_{VV}}\cos\delta\Big).
\end{gather}

Determinant of the system  of two equations  (\ref{modif}) equals
\begin{equation}
 \label{Det-new}
  D=\begin{vmatrix} (\sqrt{w_{HH}}+\sqrt{w_{VV}}\cos\delta)& \sqrt{w_{VV}}\sin\delta\\
 \,\\
 -\sqrt{w_{VV}}\sin\delta & \sqrt{w_{HH}}+\sqrt{w_{VV}}\cos\delta \end{vmatrix}=
 w_{HH}+w_{VV}+2\sqrt{w_{HH}w_{VV}}\cos\delta.
 \end{equation}
As long as  $D\neq 0$, Equations  (\ref{modif}) have the single-valued solutions
\begin{equation}
 \label{cos+sin}
 \cos\varphi_1=\frac{1}{ D}
 \begin{vmatrix}
 F & \sqrt{w_{VV}}\sin\delta\\
 \,\\
 F^{\lambda/8,\pi/4} & \sqrt{w_{HH}}+\sqrt{w_{VV}}\cos\delta
 \end{vmatrix},\quad
 \sin\varphi_1=
 \begin{vmatrix}
 \sqrt{w_{HH}}+\sqrt{w_{VV}}\cos\delta & F\\
  \,\\
 -\sqrt{w_{VV}}\cos\delta & F^{\lambda/8,\pi/4}
 \end{vmatrix}.
 \end{equation}
With $\cos\varphi_1$ and $\cos\varphi_1$ known, both phases $\varphi_1$ and $\varphi_3=\varphi_1+\delta$ are uniquely determined.

On the other hand, as can be found easily from Equation (ref{Det-ne}), the determinant $D$ turns zero at $w_{HH}=w_{VV}$ and $\delta=\pi$, i.e. at $C_1=-C_3$, and this is a special case, discussed below separately in the following Section.

 \section{A special case of the qutrit with  $C_1=-C_3$ (Step 5).}

 In this case the state vector of the qutrit has the form
 \begin{equation}
 \label{special}
 \ket{\Psi_{sp}}=|C_1|e^{i\varphi_1}(\ket{2_H}-\ket{2_V})+C_2\ket{1_H,1_V},
\end{equation}
where $|C_1|^2=w_{HH}$ and $C_2^2=w_{HV}^{tot}=2w_{HV}$  are assumed to be known, and the only unknown quantity is the phase $\varphi_1$ or $\sin\varphi_1$ and $\cos\varphi_1$.
In this case, it appears convenient to perform measurements with the qutrit preliminary transformed by the $\lambda/8$ phase plat with orientations of its optical axis: at $\alpha=0$ and at $\alpha=45^\circ$. In both cases it's sufficient to count only photons with horizontal polarization. In the case of the horizontally oriented optical axis, in accordance with the first expression in Equations (\ref{B-1-2-3-lambda/8}) and with the equality $C_3=-C_1$ taken into account the searched probability is given by \begin{equation}
 \label{wHH-sp-l/8}
 w_{HH}^{\lambda/8}=|B_1^{\lambda/8}|^2=\left|C_1\frac{(1-i)}{2}+
 C_2\frac{\sqrt{i}}{\sqrt{2}}\right|^2=\frac{w_{HH}}{2}+w_{HV}-
 \sqrt{\frac{w_{HH}w_{VV}}{2}}\,\sin\varphi_1.
\end{equation}
In the case of the wave plate's optical axis turned for $45^\circ$, the same probability is determined by the first of tree equations (\ref{C-45}) which is significantly simplified at $C_1=-C_3$:

\begin{equation}
\label{wHH-sp-l/8-45}
 w_{HH}^{\lambda/8,\pi/4}=
 \left|B_1^{\lambda/8,\pi/4}\right|^2=\left|C_1+\frac{C_2}{2}\right|^2=
 w_{HH}+\frac{w_{HV}}{2}+\frac{\sqrt{w_{HH}w_{HV}}}{2}\,\cos\varphi_1.
\end{equation}
Two equations (\ref{wHH-sp-l/8}) and (\ref{wHH-sp-l/8-45}) together determine both $\sin\varphi_1$ and $\cos\varphi_1$ as well as the phase $\varphi_1$ itself. As for the phase $\varphi_3$, in the special case under consideration it equals either  $\varphi_3=\varphi_1+\pi$ or $\varphi_3=\varphi_1-\pi$, and the choice between these two values has to be done in favor of that one which belongs to the interval $[-\pi,\pi]$.

\section{Conclusion}

 A scheme is suggested and described  in details for the direct, complete and single-valued recovery of all parameters of biphoton polarization qutrits.

\section*{Acknowledgement}
The work was partially supported by the Russian Scientific Foundation, the grant No. 22-12-00396 (https://rscf.ru/project/22-12-00396/).
Authors thank Dr. K.G. Katamadze for fruitful discussions.

\end{document}